# Microwave-clock timescale with instability on order of $10^{-17}$

Steven Peil, Thomas B. Swanson, James Hanssen and Jennifer Taylor

United States Naval Observatory, 3450 Massachusetts Ave. NW, Washington, DC 20392 USA

steven.peil@usno.navy.mil

**Abstract**. The fundamental limits of atomic fountains as operational clocks are considered. Four rubidium fountains in operation at the U.S. Naval Observatory for over 5.5 years have demonstrated unprecedented long-term stability for continuously running clocks [1,2]. With only these rubidium fountains, a post-processed timescale can be created that demonstrates superior long-term performance to any individual clock by compensating for occasional frequency steps. By comparing to the world's primary standards we demonstrate instability of this rubidium fountain timescale reaching the mid $10^{-17}$'s and zero drift at the level of $1.3 \times 10^{-19}$/day. We discuss fundamental limits due to common mode behaviour or individual fountain performance that cannot be corrected.

## 1. Rubidium fountains at USNO

The world's atomic time relies on the availability of hundreds of operational (continuously running) atomic clocks. Clock data are collected by the BIPM, which assigns an independent weight to each device based on its performance and produces free-running (EAL) and calibrated (TAI, UTC) timescales. Frequency calibrations are provided by primary standards that serve as an absolute reference. Presently, these consist of cesium fountains operated at about a dozen metrology labs worldwide, with the expectation that optical standards may soon contribute.

Four rubidium fountains at the U.S. Naval Observatory (USNO) in Washington, DC have participated in the ensemble of operational clocks used to generate these atomic timescales for the past 4.5 years [3]. These fountains are the first cold-atom clocks to be used in continuous operation, in the spirit of hydrogen masers or commercial cesium beams, and currently they are the only cold-atom systems that report to the BIPM as continuously running clocks. As of MJD 57719, the USNO fountains contributed to and received maximum weight in each Circular T [4] since MJD 56074, a total of 55 consecutive monthly reports (Fig. 1).

While masers and commercial cesium clocks are mature technologies that have been studied extensively [6–11], we are still learning about the performance of the rubidium fountains. In fact, the limit of these systems as operational clocks is an ongoing investigation. Here we report on the performance of a post-processed timescale generated with the four USNO rubidium fountains over the past 5.5 years and we compare this to the cesium fountains used to calibrate EAL. We demonstrate the gain in long-term performance associated with identifying frequency changes, and we discuss what the ultimate limits to the performance of this timescale might be.

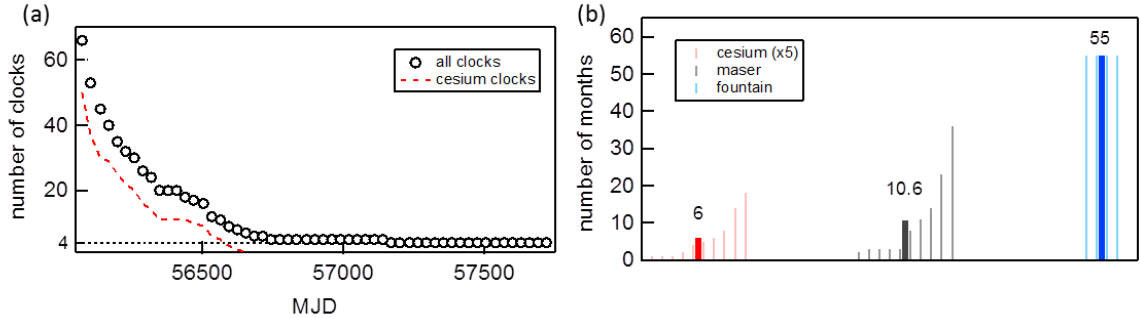

Figure 1 (color online) (a) As of MJD 57719, the number of clocks receiving maximum weight for each Circular T reporting period consecutively since MJD 56074, versus MJD. Some of the trend may reflect a change to the TAI algorithm applied in MJD 56654 [5]. The maximum weight varies each month, ranging from ~0.7% to over 1% for this time period. (b) For each type of clock (hydrogen maser, commercial cesium clock, rubidium fountain), the number of consecutive months since MJD 56074 that a clock of that type received maximum weight. The cesium data points (light red bars [left]) each represent an average of 5 clocks. The bold lines and the numeric label give average values. The numbers for masers and cesium clocks are fixed.

## 2. Post-processed rubidium fountain timescale

The behaviour of the individual rubidium fountains in operation at USNO has been reported elsewhere [1,2]; it is dominated by white-frequency noise, with all four fountains showing intervals of time over which the instability integrates below $10^{-16}$. The non-stationary behaviour that is observed in pairwise fountain comparisons is consistent with rare occurrences of discrete frequency steps in an individual device. A step is identified as a change in slope of the relative phase between one fountain and the other three. Because these jumps in frequency are rare and because the fountains achieve high uptime (over 99.5% on average), the clock which experienced a frequency change can be unambiguously identified. Over more than five years of continuous operation, we have observed an average of 1.4 frequency changes per fountain per year. The steps can be in either direction, with an average magnitude of $6.0\times10^{-16}$ and an average value of $-1.4\times10^{-17}$. There is not a strong enough correlation between the frequency steps and external events to assign a particular cause, but several steps seem to have occurred around the time of some type of user intervention, such as laser replacement.

According to pair-wise comparisons, there is no other observable source of non-Gaussian noise in the clocks. In Fig. 2 we show a sigma-tau plot for two of the fountains, NRF4 vs NRF5, after correcting for several frequency steps in each device (using all four clocks for the determination of frequency changes). The sigma-tau plot does not rise above the $1/\sqrt{\tau}$ line in the long term, out to averaging times of over 3 years, indicating that there is no relative source of flicker or random-walk that manifests itself as anything other than these rare frequency steps. Furthermore, no drift has been removed, and none has been observed in any of the fountains that cannot be attributed to discrete frequency steps. (The fact that the instability data points go below the $1/\sqrt{\tau}$ line for large $\tau$ does not reflect physical behaviour; our simple recharacterization process tends to "whiten" the relative phase records, resulting in a $1/\tau$ trend for times longer than the average time between steps.)

Once a frequency step has been identified, the corresponding clock's frequency record can be post-processed and averaged with the record of the remaining fountains. We use a simple average with equal weight for each fountain. The resulting timescale displays superior long-term stability to any individual, free-running device [1,2]. The short-term stability is improved by a factor of 2 simply by averaging the white-frequency noise, as demonstrated previously with 2 years of data [1]. That shorter data set did not show any improvement in long-term performance of the re-characterized rubidium fountain timescale over the raw timescale formed without making any frequency corrections.

There are various limitations that can be expected for this type of timescale. In terms of individual clock behaviour, some frequency changes may not be correctable. For a clock comparison governed by white-frequency noise with a 1s Allan deviation of $\sigma$, the averaging time that is required to achieve unity signal-to-noise on the detection of a frequency step of size $\Delta$ is of order $(\sigma/\Delta)^2$. So there may be frequency variations in a clock that occur with a particular size and at a particular (average) rate that would not be identifiable as discrete jumps.

Furthermore, when averaging the fountains to produce a timescale, there is no way to remove common mode trends. This could apply to frequency changes, but more likely is relevant for potential drifts. As an example, it was discovered by comparing to the rubidium fountains that the USNO cesium timescale, formed using only commercial cesium clocks, had a drift of $2\times10^{-17}$/day over a particular epoch [13]. That low level of drift for clocks of a common type is not surprising, yet it could not be determined without resorting to an external reference for comparison. Thus, outside comparisons are necessary to assess the performance of the re-characterized fountain timescale.

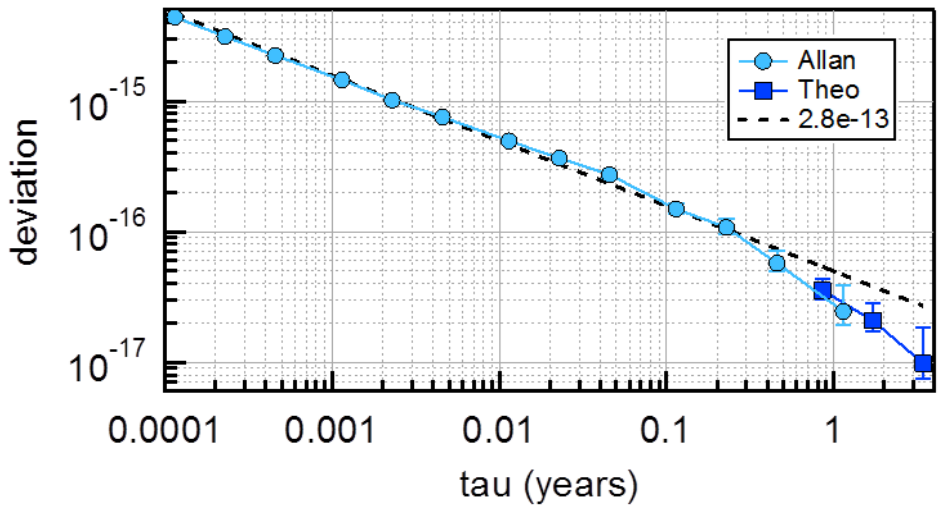

Figure 2 (color online) Allan and Theo [12] deviations for NRF4 vs NRF5 over 5.5 years after removing several frequency steps (using all four fountains). No behaviour outside of these steps produces long-term instability that lies above the dashed $1/\sqrt{\tau}$ line. The fact that points fall below the $1/\sqrt{\tau}$ line is non-physical, resulting from re-characterizing using the phase records [2]; our simple technique ends up "whitening" the data.

## 3. Comparison to cesium-fountain primary standards

Arguably the best long-term frequency reference that we can use to assess the rubidium fountain timescale is the average frequency of the cesium-fountain primary standards reporting to the BIPM. This frequency is published monthly in the BIPM's Circular T and can be compared against a monthly average of our fountain timescale.

The rubidium fountains are measured regularly with a low-noise measurement system against USNO's master reference clock, and we evaluate monthly averages of the fountain ensemble against the master clock, $\nu_{USNO}-\nu_{Rb}$. Using the published values of UTC-UTC(USNO) in the Circular T, we can determine the monthly average of $\nu_{TAI}-\nu_{USNO}$ (since UTC and TAI have the same frequency), allowing us to determine $\nu_{TAI}-\nu_{Rb}$. These values can then be adjusted by $\nu_{TAI}-\nu_{Cs}$, the frequency difference between TAI and the cesium-fountain primary standards, which is also published monthly in the Circular T. This method has been used in the past [1], and we update it here for 5.5 years of data. The frequency comparison and the corresponding sigma-tau plot are shown in Fig. 3.

The rubidium-cesium fountain comparison can also be made using the values of the rubidium fountain frequencies published in the Circular T each month. These are derived by the BIPM from data provided by USNO, a combination of time transfer of UTC(USNO) and reports of individual clock phases. These published frequencies are referenced to TAI, so we can apply the re-characterization to correct for frequency jumps and make the same adjustment as above to arrive at

$\nu_{Cs}-\nu_{Rb}$. This method gives results that are consistent with the previous technique, but with fewer data points due to the delay in reporting the fountains [3].

The stability plot, Fig. 3(b), includes a dashed black line corresponding to a white frequency noise level of $10^{-13}$, a lower limit for the rubidium-fountain timescale, and a dashed red line serving as a reference level for a drift rate of $5\times10^{-20}$/day (using the upper limit on the last stability point as a guide). The instability data show additional white frequency noise above the level expected from the rubidium timescale, but the trend shows no deviation from $1/\sqrt{\tau}$ behaviour in the long term, going out to averaging times of several years and instability levels in the $10^{-17}$'s.

Thus, there is no evidence of any long-term non-stationary behaviour in the fountains that is not removed in the post-processing associated with removing rare, discrete frequency steps. Furthermore, there is no evidence of common mode drift with respect to the cesium primary standards. A linear fit, shown in Fig. 3(a), shows no relative drift between the two timescales at the level of $1.3\times10^{-19}$/day. The (absolute) uncertainty on the fit is $\pm1.07\times10^{-19}$/day=$\pm3.9\times10^{-17}$/year, to our knowledge, the lowest uncertainty on the drift between frequencies of two different atomic species, even including optical transitions [14,15]. References [14] and [15] describe frequency measurements with accuracy evaluations. While we do not specifically carry out accuracy evaluations, we gain confidence that the *temporal* behaviour of the re-characterized ensemble frequency is free from systematic perturbations, at the level of effects we can model and at the level that comparison to the cesium standards indicates.

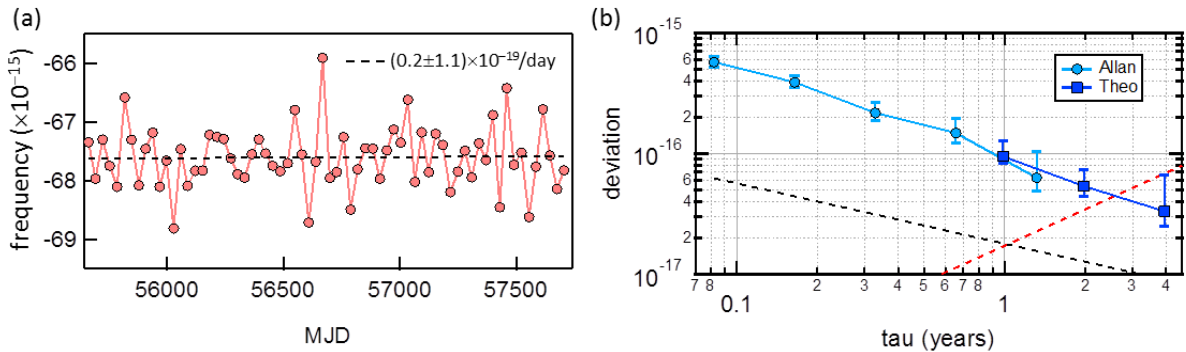

Figure 3 (color online) (a) Frequency difference between the USNO rubidium-fountain timescale and the weighted average frequency of the cesium fountains reporting to the BIPM, $\nu_{Cs}-\nu_{Rb}$. The data are derived using local measurements of the rubidium fountains with the USNO master clock and adjusting with frequency differences published in the Circular T. The dashed line is a linear fit, showing zero drift between the two sets of clocks at a level of $1.3\times10^{-19}$/day. (b) Sigma-tau plot corresponding to this frequency record. Dashed lines are white-frequency noise (black) and drift (red) reference levels. The averaging time "tau" ranges from .08 to ~4 years.

## 4. Fundamental limits

We can consider what the fundamental limits of the rubidium fountain timescale might be, at least in comparison to the cesium (or other future) primary standards. Known and potential sources of frequency biases in an atomic fountain have been analysed for many different systems; a succinct review can be found in Ref. [16]. Considering limits to stability from possible variations in these frequency biases by analysing fluctuations in the parameters that drive them can help us predict the fundamental limits to stability of our fountain timescale.

### 4.1. Individual fountains

As discussed in Section 2, we expect a limitation due to frequency steps in an individual clock of a size and at a rate that make them undetectable using a collection of other clocks. We suspect that some of the frequency steps that we do correct for may be due to instabilities in the microwave drive, possibly related to stray coupling to RF signals, while others are suggestive of changes in an AC Stark shift due to stray light, seemingly correlated with user intervention. These same issues may cause smaller frequency changes that are not detectable and therefore not corrected with our re-characterization procedure. Along with other technical sources of instability [16], it is difficult to model the size of such potential steps. Other possible sources of small frequency steps that would not get removed in the timescale generation are associated with well understood models, such as number or density fluctuations driving cold collisions [17], magnetic-field fluctuations driving the second-order Zeeman shift, and certain cavity effects (such as distributed cavity phase shift, microwave leakage, and line pulling [16]), and we can attempt to estimate the impact of these.

We measure the variations in the number of atoms launched over time by recording the detected fluorescence on each fountain cycle. We can convert these variations into frequency fluctuations via the density dependence of the collision shift, allowing us to estimate that cold collisions introduce flicker in an individual fountain at a level of order mid $10^{-18}$'s. We measure the average magnetic field over the atomic trajectory every several hours by measuring a field-sensitive atomic transition. The frequency fluctuations due to these magnetic-field variations (from changes in the quadratic Zeeman shift) produce a flicker level below $10^{-18}$. In terms of cavity related shifts, we can model the effect of temperature fluctuations on line pulling and distributed cavity phase shifts (although variations in distributed cavity phase could also result from changes in the angle of the atomic trajectory with respect to orientation of the microwave field). However, the four fountains were built with two different cavity designs, including detuning from atomic resonance and Q, with a pair of fountains incorporating each design. Therefore, this type of cavity related instability would arise partly as common mode behavior; we do not try to estimate the effect of these cavity issues here. Instabilities in individual fountains due to effects that are straightforward to model appear to impact stability at a level below $10^{-17}$.

*4.2. Common mode behavior*

A second potential limit to the performance of the rubidium fountain timescale is behaviour that is common in all of the fountains. This could be technical in nature, such as common environmental (temperature, humidity or atmospheric pressure) variations in the building where the fountains are housed, or more fundamental, such as variations in the Earth's magnetic field or the gravitational potential where the fountains reside. Because we do not carry out accuracy evaluations, biases are not corrected for and changes in their values would appear as an instability in a comparison against primary frequency standards.

To get a feel for the impact of some of these common mode effects, we can convert the limit at which the drift of the rubidium timescale is shown to be zero, $1.3\times10^{-19}$/day, to a limit on drift rate of different common-mode biases, assuming that we can treat each independently. If we model just the effect of blackbody radiation, and neglect the effect of associated mechanical changes, the data limit the average drift over the past four years of the temperature of the clock rooms to below 2 mK/day. Such an effect would be caused by a drift in the reading of the sensors that measure the room temperature, which are redundant and which are recalibrated every several years. Using a modelled shielding factor of our magnetic shield set of $10^5$, the limit on frequency drift translates to a limit on the average drift of the Earth's local (Washington, DC) magnetic field of 85 µG/day. From the gravitational redshift, the frequency drift limit indicates that the drift in elevation of Washington, DC with respect to the geoid is zero at the level of 1.3 mm/day over 5.5 years.

The limits on the average drift rates of these physical parameters are all reasonable, and in fact not very stringent. We can consider a more aggressive method of assigning a limit to the frequency drift of our timescale, say by using the upper limit of the last Theo point calculated, giving $5\times10^{-20}$/day (Fig. 3(b)), though this is still not enough to be able to say something interesting about the potential drivers behind common mode behavior. For example, this more stringent limit is still an order of magnitude above what would be needed to be able to see the maximum effect of glacial isostatic adjustment (GIA), the slow rebound of the Earth's crust in certain locations following glacial retreat after the last ice age [18]. Improved time transfer and state-of-the-art LOs in our fountains could push the white-frequency noise level of a comparison between a fountain timescale and the cesium fountains low enough to perhaps make observing such effects possible.

Aside from drift, realistic fluctuations in any of these common mode effects are unlikely to impose a flicker floor at the $10^{-17}$ level or above.

## 5. Searches for new physics

Frequency differences between different species of atomic clocks have been used to search for new physics by searching for certain temporal signatures, such as drifts or oscillations. Limits on the drift in the frequency difference can be used to put limits on possible drifts in the values of fundamental constants [14], constraining possible violations of local position invariance (LPI), while limits on harmonic oscillations can be used to constrain scalar field models of dark matter [19,20].

Some of these past measurements have been carried out using rubidium versus cesium, including work we carried out previously using rubidium fountains and commercial cesium beams [13]. We can use the frequency comparison above (Fig. 3(a)), between our rubidium ensemble and the cesium fountain primary standards, to add constraints to or improve upon some of these measurements. For example, searching for an annual oscillation in phase with the Earth's solar orbit provides a test of LPI and a method of constraining the gravitational dependence of fundamental constants. This test arises from the fact that, according to the equivalence principle, the gravitational redshift of a clock is independent of the type of clock, and the varying gravitational potential provided by the Earth's elliptical orbit around the Sun provides a driving term for determining whether one type of clock responds with a different redshift than another. A sinusoidal fit to the data in Fig. 3(a) gives a smaller uncertainty on an anomalous differential redshift of these two species of clock than any other comparison published to date, $\beta_{Cs}-\beta_{Rb}=(7.6\pm5.5)\times10^{-7}$, where $\beta$ characterizes the degree of LPI violation manifest in a gravitational redshift measurement [13]. We anticipate continuing to add to our frequency comparison record and may consider contributing to some of the other constraints being placed on physics beyond the standard model.

## 6. Real-time performance

One obvious drawback to the rubidium-fountain timescale is the fact that it is post-processed and therefore is not available in real time. We can look at the performance of a "real-time" timescale [21] created by simply averaging the four rubidium fountains, again using the cesium fountains reporting to the BIPM as a reference. In this case, we use the same method as discussed above to derive the frequency comparison, but now we do not apply any frequency corrections to the rubidium fountains' frequency records. The sigma-tau plot of the resulting timescale is shown in Fig. 4. By comparison with Fig. 3(b), the stability plot shows that the frequency corrections we apply to derive our post-processed timescale do indeed make an improvement over the raw timescale, but that even without these corrections the four fountains produce a timescale that integrates below $2\times10^{-16}$, where there is evidence of flicker behavior. This is impressive performance for a real-time timescale, requiring no re-characterization of clock behavior to reach these levels.

We conclude the discussion of our fountain timescale with a brief remark on the implications for assessing the relative performance of atomic and astrophysical clocks. In Ref. [22], a re-analysis of pulsar (and white dwarf) timing data showed that atomic clocks exhibit superior stability to those astrophysical systems out to averaging times of about 2 years, where the best clock comparisons reached instabilities slightly better than $10^{-15}$. That is roughly the same as the stability measured for pulse trains from the best pulsar at about 10 years of averaging. The availability of atomic clock-comparison data was limited at the longest averaging times analysed since most of the highest performing clocks do not operate continuously for years at a time. The long-term stability of $10^{-16}$ that we demonstrate here (Fig. 4) is explicit evidence of superior stability of atomic clocks compared to pulsars at averaging times of several years. In fact, the Allan variance for the pulsar timing data is only analysed after a drift in the pulse period is removed. This suggests that our re-characterized timescale shown in Fig. 3(b) may be the more appropriate reference to compare to pulsar stability, and that comparison demonstrates an even greater improvement in stability of atomic clocks compared to astrophysical ones, roughly 2 orders of magnitude difference at 2 years of averaging.

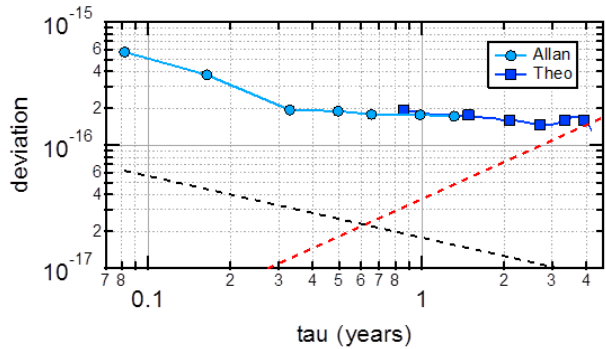

Figure 4 (color online) Allan and Theo deviations for free-running fountain timescale, measured against the cesium fountain primary standards. Dashed lines are white-frequency noise (black) and drift (red) reference levels, as in Fig. 3(b). The averaging time "tau" ranges from .08 to ~4 years.

## 7. Conclusions

Using inter-comparisons among four rubidium fountains at USNO, we can identify and remove rare frequency steps and create a post-processed timescale that shows excellent long-term stability compared to the world's cesium-fountain primary standards. Straightforward modelling of sources of instability indicate there are no obvious fundamental limits to the post-processed timescale at the $10^{-17}$ level. In addition to uses for local timing, the timescale may be applicable as a flywheel for comparing different primary frequency standards reporting to the BIPM as well as for adding constraints to models of new physics beyond the standard model. The achievement of order $10^{-16}$ frequency stability for a real-time atomic-clock timescale over years of averaging further demonstrates the superiority of atomic clocks compared to astrophysical clocks that was summarized in Ref [22].

## 8. Acknowledgements

We thank Chris Ekstrom, who led the rubidium fountain effort at USNO for many years, and Demetrios Matsakis for discussions of data analysis.